\documentclass[reprint,floatfix,showpacs,showkeys,aps,prb]{revtex4-2}
\usepackage[T1]{fontenc}
\usepackage[utf8]{inputenc}
\usepackage[english]{babel}
\usepackage{graphicx}
\usepackage{times}
\usepackage{mathptmx}
\usepackage{eurosym}
\usepackage{amssymb,amsmath,amsfonts}
\usepackage{bm}
\usepackage{latexsym}
\usepackage{subfigure}
\usepackage{float}
\usepackage{url}
\usepackage{hyperref}
\usepackage{tensor}
\usepackage{xfrac}
\usepackage{multirow}
\usepackage{tabulary}
\usepackage{fancyhdr}
\usepackage{MnSymbol}
\usepackage{placeins}
\usepackage{booktabs}
\usepackage[flushleft]{threeparttable}
\usepackage{lineno}
\usepackage{balance}
\usepackage[usenames]{color}
\usepackage{lipsum}
\usepackage{mathtools,cuted}
\usepackage{setspace}
\usepackage{chngcntr}

\begin{document}

\title{A Boson-Fermion theory that goes beyond the BCS approximations for superconductors}

\author{I. Ch\'{a}vez}
\author{P. Salas}
\author{M.A. Sol\'{i}s}
\address{Instituto de F\'{i}sica, Universidad Nacional Aut\'{o}noma de M\'{e}xico,
Apdo. Postal 20-364, 01000 Mexico City, MEXICO}
\author{M.~de~Llano}
\address{Instituto de Investigaciones en Materiales, Universidad Nacional Aut\'onoma de M\'exico, 04510 Mexico City, MEXICO}
	
\date{\today}

\begin{abstract}
A detailed analysis is given of the effects of common and recurring approximations used in conventional superconductivity theories on the condensation energy values, whose magnitudes are notoriously smaller than those of other energies as the superconducting energy gap and the chemical potential. These approximations come from using the density of states $N(\epsilon)$ and the chemical potential $\mu(T)$ either constant or temperature-dependent, respectively. We use these approximations, a total of three, to calculate the critical temperature $T_c$, the superconductor energy gap $\Delta(T)$, the chemical potential $\mu(T)$ and the thermodynamic potential $\Omega(T)$ which are needed to obtain the condensation energy, and compare them with the \textit{exact} case, i.e., where no approximations are used. To do this, we use a ternary Boson-Fermion theory of superconductivity composed of unbound electrons (or holes) as fermions plus two-electron and two-hole Cooper pairs, both as bosons. Although all these approximations lead to reasonable values of $T_c$ and $\Delta(T)$, the resulting thermodynamic and chemical potentials are quite different, so that the condensation energy value could be incorrect. However, when $N(\epsilon)$ and $\mu(T)$ variables are used, together with a correct physical interpretation of the condensation energy as the sum of the thermodynamic and chemical potential differences, it leads to a better agreement with reported experimental data, compared to the one obtained when taking them as constants, particularly so for low temperatures.
\end{abstract}

\maketitle

\section{Introduction}

In 1957 J. Bardeen, L. Cooper and J. Schrieffer \cite{BCS} formulated what is now known as the BCS theory of superconductivity, which microscopically addresses the current-without-electrical-resistance in a superconductor (SC) through the Cooper pairing mechanism \cite{cooper}. These electron pairs are bound as long as the energy difference between them is less than the energy of a phonon ($\hbar\omega_D$), a state in which the superconducting phase can be formed when the attractive interaction surpasses the coulombian repulsion. This minimum excitation energy is the energy gap $\Delta(T)$, which at zero-absolute temperature is $\Delta_0 = 1.75 k_BT_c$, where $k_B$ is the Boltzmann constant and $T_c$ the critical temperature. 

BCS theory depends on three parameters:  the density of states (DOS) at the Fermi level $N(\epsilon_F)$,  the average velocity of electrons at the Fermi surface  $v_0$ and the electron-phonon interaction. Thus, in the BCS theory the basic problem reduces to calculate the ground state and excited states of the fermion system, which interact via a two-body potential. Although it is not the purpose here to make any revision on the BCS work, we analyze and point out some aspects of the theory, and take some of its equations to compare with our results. 

BCS starts with a reduced problem which only includes the configuration of the states occupied by pairs, where the total number of particles of the system is fixed and formed by electrons when $k > k_F$, holes with $k < k_F$ and pairs formed via a negative interaction $V_{kk'}$. The BCS Hamiltonian considers  pairs with zero net momentum and neglects the ones with total momentum different from zero, and supposes a symmetric distribution function between electrons and holes with respect to the Fermi surface. In order to find the ground state relative to the Fermi sea, the theory uses a Hartree-like wave function as a variational approximation.

So, for states within a range $|\epsilon_k| < \hbar \omega_D$ one obtains the condition on $\Delta_0$
\begin{equation}
\frac{1}{V} = \sum_{k} \frac{1}{2(\xi_k^2 + \Delta_0^2 )^{1/2}} \label{BCS-Int}
\end{equation}
where $\xi_k$ is a single particle energy with respect to the Fermi energy and $V$ is a constant average matrix element which is related to the matrix element for the electron-phonon interaction. Replacing the sum by an integral gives
\begin{equation}
\frac{1}{N(\epsilon_F) V} = \int\limits_{0}^{\hbar\omega_D} \frac{d\xi_k}{(\xi_k^2 + \Delta_0^2 )^{1/2}} \label{BCS-lambda}
\end{equation}
where $N(\epsilon_F)V = \lambda_{BCS}$ is the dimensionless coupling constant of the BCS theory. Solving for $\Delta_0$ in the weak coupling limit gives
\begin{equation}
\Delta_0 = \hbar\omega_D\; \sinh^{-1} (-1/\lambda_{BCS}).
\end{equation}
Using the relation between the energy of the ground state  and $k_BT_c$ leads to the estimation 
\begin{equation}
k_BT_c \sim \hbar\omega_D \exp(-1/\lambda_{BCS}).
\end{equation}
Thus the energy gap and the critical temperature are related to the electron-phonon interaction \cite{bardeen55, morel-anderson} and to the DOS.

In the BCS theory the temperature dependence of the chemical potential has been omitted; an approximation that has been taken not only by BCS but also by several solid state academics \cite{ashcroft}, particularly when studying metals, where $\mu$ is assumed equal to $E_F$  all the way from $T = 0$ to room temperature. Although it is essential to keep in mind that the chemical potential is different from its zero temperature value, here we show that it is important to focus on that inequality. Thus, we focus on the last affirmation and on how the assumption of taking DOS constant in the normal free energy leads to innacurate calculations for the condensation energy of the superconductors, even though $\Delta_0$ and $k_BT_c$ are well determined. Furthermore, we show that the order of magnitude of the difference between the superconducting and normal chemical potentials is of the same order of magnitude as the difference between the superconducting and normal thermodynamic potentials and that both quantities are essential for the correct calculation of the Helmholtz free energy and the condensation energy.

In section II we recall the Boson-Fermion (BF) theoretical background \cite{GBEC1,GBEC2} and the main framework equations with the following approximations i) DOS constant and $\mu = E_F$ constant, as assumed by BCS, ii) DOS constant and $\mu$ variable, iii) DOS variable and $\mu = E_F$; and we compare the results with iv) where both  DOS and $\mu(T)$ are energy and temperature dependent,  which we call the \textit{exact} case. In section III we calculate the Helmholtz free energy and condensation energy from the BF mixture. We show the importance of carefully  and consistently taking the above mentioned approaches and we compare our results with reported data for superconducting aluminum. Finally, we offer our conclusions.

\section{Boson-Fermion formalism}

The generalized Bose-Einstein condensation (GBEC) theory starts from an ideal BF ternary gas consisting of unbound electrons/holes (fermions) with two-electron Cooper pairs (2eCPs) and two-hole Cooper pairs (2hCPs) both as bosons, with very particular BF vertex interactions. It is described \cite{GBEC1, GBEC2} by the Hamiltonian $H = H_0 + H_{int}$ where $H_0$ is an ideal ternary gas and $H_{int}$ contains the BF vertex interactions.

We take $\mathbf{K}\equiv \mathbf{k}_{1}+\mathbf{k}_{2}$ as the center-of-mass momentum (CMM) wavevector of two fermions, where $K \equiv \vert \mathbf{K}\vert $, and $\mathbf{k}\equiv \frac{1}{2}(\mathbf{k}_{1}-\mathbf{k}_{2})$ is their relative wavevector. Ignoring composite bosons with $K\neq 0$ in $H_{int}$---but \textit{not} in $H_{0}$, as assumed in BCS theory--- one can consider a simpler \textit{reduced} $H_{red}$. If one applies the Bogoliubov recipe of replacing the zero-$K$ creation operators $b_{\mathbf{0}}^{\dagger}$ and $c_{\mathbf{0}}^{\dagger }$ for the 2e/2hCP bosons by  $\sqrt{N_{0}}$ and $\sqrt{M_{0}}$, with $N_{0}$ and $M_{0}$ the numbers of 2e/2hCP $K=0 $ bosons, and using the Bogoliubov-Valatin transformation \cite{bogo58, valatin} allows for the exact diagonalization of the dynamical operator $\hat{H}_{red}-\mu \hat{N}$ \cite{GBEC3}, with $\hat{N}$ the total-electron-number operator and $\mu $ a Lagrange multiplier. 

The thermodynamic  potential of the grand-canonical statistical ensemble for this ternary BF mixture \cite{GBEC1, GBEC2} is $\Omega (T,L^{3},\mu ,N_{0},M_{0}) =-k_{B}T\ln \left[\mathrm{Tr}(\exp \{-\beta (\hat{H}_{red}-\mu \hat{N})\})\right] $, where \textrm{Tr} means \textquotedblleft trace,\textquotedblright\ $L^{3}$ is the
3D system volume, $\mu$ is the BF chemical potential and $\beta \equiv 1/k_{B}T$. Specifically, the thermodynamic grand potential becomes

\begin{flalign}
\frac{\Omega}{L^{3}} &=  \int_{0}^{\infty }d\epsilon_k~ N(\epsilon_k )[\epsilon_k -\mu -E(\epsilon_k )]  \label{omega}
\\
&- 2k_{B}T\int_{0}^{\infty }d\epsilon_k~ N(\epsilon_k )
\ln \left( 1+\exp [- \tfrac{E(\epsilon_k )]}{k_B\,T}] \right)  \notag \\
&+  [E_{+}(0)-2\mu ]n_{0} \notag \\
&+ k_{B}T\int_{0^{+}}^{\infty }d\varepsilon_K~ M(\varepsilon_K )
\ln \left( 1-\exp [-\tfrac{E_{+}(0)+\varepsilon_K -2\mu}{k_B\,T} ]\right)  \notag \\
&+ [2\mu-E_{-}(0)]m_{0} \notag \\
&+ k_{B}T\int_{0^{+}}^{\infty }d\varepsilon_K~ M(\varepsilon_K )
\ln \left( 1-\exp [- \tfrac{2\mu -E_{-}(0)+\varepsilon_K}{k_B\,T}]\right) \notag
\end{flalign}
\noindent where $n_0 \equiv N_0/L^3$, $m_0 \equiv M_0/L^3$ is the number density of 2eCPs and 2hCPs for $K=0$, $N(\epsilon_k)$ the fermionic DOS, $M(\varepsilon_K)$ the bosonic DOS, with $\epsilon_k = \hbar^2 k^2/2m$  the fermion energy-dispersion relation and $\varepsilon_K = E_{\pm}(0) \pm \hbar^2 K^2/2(2m) $ the boson energy-dispersion relation and $E_{\pm}(0) = 2E_f \pm \delta\epsilon$ the boson intrinsic  energies with $K=0$, with $E_f$ a pseudo-Fermi energy of the unbound fermions and $\delta\epsilon$ the BF interaction range energy to bind CPs. Here, $E(\epsilon_k) = \sqrt{(\epsilon_k-\mu)^2 + \Delta^2(T)} $  and $\Delta(T) = f_{+}\sqrt{n_0(T)} + f_{-}\sqrt{m_{0}(T)}$ the energy gap, with $f_{\pm}$ the BF vertex interaction functions as originally defined in Refs. \cite{GBEC1, GBEC2}.
To find the Helmholtz free energy per unit volume, we use the equation
\begin{equation}
F(T,\mu ,N_{0},M_{0}) =  \frac{\Omega(T,\mu ,N_{0},M_{0})}{L^3} + \frac{N \mu(T)}{L^3}, \label{Eq-Helmholtz}
\end{equation}
and for the condensation energy per unit volume
\begin{eqnarray}
E_{cond} &=& F_{\mathtt{s}} - F_{\mathtt{n}}
= \left[ \Omega_{\mathtt{s}}- \Omega_{\mathtt{n}} \right]/L^3 + n \left[ \mu_{\mathtt{s}} - \mu_{\mathtt{n}} \right] \label{Eq-Econd}
\end{eqnarray}
where subscripts $\mathtt{s,n}$ denote superconductor (SC) and normal states, respectively. Taking the 50--50 proportions between 2e/2hCPs, i.e., $n_0(T) = m_0(T)$ and $n_{B+}(T) = m_{B+}(T)$ the SC Helmholtz free energy $F _{\mathtt{s}}$ of this ternary BF theory is
\begin{widetext}
\begin{small}
\begin{eqnarray} 
F_{\mathtt{s}} &=& 2\, \delta\epsilon\, n_{0}(T) + \int\limits_{E_f-\delta\epsilon}^{E_f+\delta\epsilon} \,d\epsilon_k~ N(\epsilon_k ) \left[\epsilon_k -\mu_{\mathtt{s}}(T) -\sqrt{(\epsilon_k-\mu_{\mathtt{s}}(T))^2 + \Delta^2(T)} \right] 
+ \left( \int\limits_{0}^{E_f-\delta\epsilon} + \int\limits_{E_f+\delta\epsilon}^{\infty} \right) \,d\epsilon_k~ N(\epsilon_k )[\epsilon_k -\mu_{\mathtt{s}}(T) - |\epsilon_k - \mu_{\mathtt{s}}(T)|] 
\\
&-& 2k_{B}T  \int\limits_{E_f-\delta\epsilon}^{E_f+\delta\epsilon} \,d\epsilon_k~ N(\epsilon_k ) \ln \left( 1+\exp [-\beta \sqrt{(\epsilon_k-\mu_{\mathtt{s}}(T))^2 + \Delta^2(T)}]\right) \notag
- \left(\int\limits_{0}^{E_f-\delta\epsilon} + \int\limits_{E_f+\delta\epsilon}^{\infty} \right) \,d\epsilon_k~ N(\epsilon_k )\ln \left( 1+\exp [-\beta
|\epsilon_k-\mu_{\mathtt{s}}(T)|]\right)
\\
&+& k_{B} T 
\int_{0^{+}}^{\infty }d\varepsilon_K~ M(\varepsilon_K )
\ln \left( 1-\exp[-\beta (E_{+}(0) +\varepsilon_K -2\mu_{\mathtt{s}}(T) )]\right) \notag
+ k_{B} T \int_{0^{+}}^{\infty }d\varepsilon_K~ M(\varepsilon_K )
\ln \left( 1-\exp[-\beta (E_{-}(0) +\varepsilon_K + 2\mu_{\mathtt{s}}(T) )]\right)
+ n~\mu_{\mathtt{s}}(T)  \label{FreeH5050}
\end{eqnarray}
\end{small}
\end{widetext}

while the normal state $F_{\mathtt{n}}$ with $\Delta(T) = 0$ is
\begin{flalign}
F_{\mathtt{n}} &=  \int\limits_{0}^{\infty} \,d\epsilon_k~ N(\epsilon_k )[\epsilon_k -\mu_{\mathtt{n}}(T) - |\epsilon_k - \mu_{\mathtt{n}}(T)|] 
\notag \\
&- 2k_{B}T  \int\limits_{0}^{\infty} \,d\epsilon_k~ N(\epsilon_k )\ln \left( 1+\exp [-\beta
|\epsilon_k - \mu_{\mathtt{n}}(T)|]\right)  \notag \\
&+ n~\mu_{\mathtt{n}}(T). \label{FreeHNormal}
\end{flalign}

In order to calculate the Helmholtz free energy and the condensation energy, we must first obtain the energy gap and the chemical potential values for the BF mixture by imposing equilibrium conditions, i.e., we  minimize the free energy by taking the first partial derivative with respect to the number of 2eCPs and 2hCP with $K=0$, namely
\begin{equation}
\left(\frac{\partial F}{\partial N_0} \right) = 0 \qquad \left(\frac{\partial F}{\partial M_0} \right) = 0 \qquad \left(\frac{\partial \Omega}{\partial \mu} \right) = -N  \label{conditions}
\end{equation}
where the last expression is the partial derivative of the grand potential with respect to the chemical potential in order to find the total number of particles. The first condition in \eqref{conditions} for 2eCPs gives
\begin{small}
\begin{align}
\begin{multlined}[b][0.9\columnwidth]
2\sqrt{n_{0}(T)}[E_{+}(0)-2\mu] = 
\\
\int\limits_{0}^{E_f + \delta\epsilon}\, d\epsilon_k~ N(\epsilon_k)
\frac{\Delta_e(T) f_{+}(\epsilon_k)}{E(\epsilon_k)} \tanh\left[\tfrac{1}{2}\beta E(\epsilon_k )\right] \label{Eq1}
\end{multlined}
\end{align}
\end{small}
while the second condition of \eqref{conditions} for 2hCPs gives
\begin{small}
\begin{align}
\begin{multlined}[b][0.9\columnwidth]
2\sqrt{m_{0}(T)}[2\mu -E_{-}(0)] = 
\\
\int\limits_{E_f - \delta\epsilon}^{0} \;d\epsilon_k~ N(\epsilon_k)
\frac{\Delta_h(T) f_{-}(\epsilon_k )}{E(\epsilon_k )} \tanh\left[ \tfrac{1}{2}\beta E(\epsilon_k )\right].  \label{Eq2}
\end{multlined}
\end{align}
\end{small}
Here $\Delta_e(T) = f_{+}\sqrt{n_0(T)}$ and $\Delta_h(T) = f_{-}\sqrt{m_0(T)}$ are the energy gaps of 2eCPs and 2hCPs, respectively. We take $f_{+} = f_{-} = f$ as a special case assuming the same interaction strength for both kinds of CPs. The number equation results from the third equation of \eqref{conditions} and implies
\begin{small}
\begin{flalign}
n &= 2n_0(T) + 2n_{B+}(T) - 2m_0(T) - 2m_{B+}(T) + n_f(T) 
\notag \\
&= 2n_0(T) + 2\int\limits_{0}^{\infty} d\varepsilon_K~ M(\varepsilon_K) \left(\frac{1}{\exp[\beta \{E_{+}(0) + \varepsilon_K-2\mu)\}]-1} \right) \notag \\
&- 2m_0(T) - 2\int\limits_{0}^{\infty} d\varepsilon_K~ M(\varepsilon_K) \left(\frac{1}{\exp[\beta \{E_{-}(0) + \varepsilon_K+2\mu)\}]-1} \right) 
\notag \\
&+ \int\limits_{0}^{\infty}d\epsilon_k~ N(\epsilon_k )\left( 1-\frac{%
\epsilon_k -\mu }{E(\epsilon_k)}\tanh \left[ \tfrac{1}{2}\beta E(\epsilon_k)%
\right] \right) \label{number}
\end{flalign}
\end{small}
where $N(\epsilon_k) = m^{3/2} \epsilon_{k}^{1/2}/2^{1/2}\pi^{2}\hbar^{3}$ is the fermionic DOS and $M(\varepsilon_K ) = 2m^{3/2}\varepsilon_{K}^{1/2}/\pi^{2}\hbar^{3}$ the bosonic DOS. The first and second terms refer to the number density of 2e/2hCPs with $K=0$, respectively, while the third and fourth terms are the excited bosonic 2e/2hCPs with $K \neq 0$, and the last term corresponds to the unbound fermions.

Taking a special case when $n_0(T) = m_0(T)$, implying that $\Delta(T)= \Delta_e(T) = \Delta_h(T)$ and $n_{B+}(T) = m_{B+}(T)$, i.e., a 50--50 proportions between 2eCPs and 2hCPs, the gap-like equation becomes
\begin{small}
\begin{align}
\begin{multlined}[t][0.8\columnwidth]
\delta\epsilon = \frac{f^{2}}{2} \int\limits_{E_f - \delta\epsilon}^{E_f + \delta\epsilon } d\epsilon_k~ N(\epsilon_k )  
\\ \times 
\bigg(
\frac{1}{\sqrt{(\epsilon_k-\mu)^2 + \Delta^2(T)}}
\tanh\left[ \frac{\sqrt{(\epsilon_k-\mu)^2 + \Delta^2(T)}}{k_B\,T} \right]
\bigg), \label{Eq5050}
\end{multlined}
\end{align}
\end{small}
and the number equation is
\begin{small}
\begin{eqnarray} \label{number5050}
n &=& \int\limits_{E_f - \delta\epsilon}^{E_f + \delta\epsilon}d\epsilon_k~ N(\epsilon_k )
\\ \notag
&\times &
\bigg( 1-\frac{\epsilon_k -\mu }{\sqrt{(\epsilon_k-\mu)^2 + \Delta^2(T)}}
\tanh \left[ \tfrac{1}{2}\beta \sqrt{(\epsilon_k-\mu)^2 + \Delta^2(T)}%
\right]
\bigg)
\\ \notag
&+&
\bigg(
\int\limits_{0}^{E_f - \delta\epsilon} + \int\limits_{E_f + \delta\epsilon}^{\infty}
\bigg)
\; d\epsilon_k~ N(\epsilon_k )
\bigg( 1-\frac{\epsilon_k -\mu }{|\epsilon_k-\mu|} \tanh \left[ \tfrac{1}{2}\beta |\epsilon_k-\mu|\right]
\bigg).
\end{eqnarray}
\end{small}

We solve Eqs. \eqref{Eq5050} and \eqref{number5050} simultaneously to find the temperature-dependent energy gap and the chemical potential for the BF mixture; both expressions are now known as the BCS-Bose crossover equations for any $T \geq 0$; the general case is obtained by solving \eqref{Eq1}, \eqref{Eq2} and \eqref{number} is the BCS-Bose crossover extended \cite{chavez17,chavez18} with 2hCPs.

\subsection{Approximations by BCS}

Historically, the BCS theory calculated the energy gap assuming both the chemical potential $\mu$ and the density of states (DOS) as constants \cite{BCS}. Here, we will present the (i) case, by assuming DOS and $\mu$ constants and will later compare our results with experimental data.

Within the BCS theory, a symmetrical distribution between electrons and holes near the Fermi energy is assumed, which  corresponds to the case when the same distribution of 2eCPs and 2hCPs is taken in the ternary BF theory, i.e., a 50--50 proportions. Since DOS is identically the same for 2eCPs and 2hCPs, and $n_0(T) = m_0(T)$, this leads to a 50--50 gap-like equation \eqref{Eq5050} due to the fact that the BF interaction function occurs  in the energy range $[E_f - \delta\epsilon, E_f + \delta\epsilon]$. Taking this symmetric distribution \cite{GBEC1} one gets
\begin{small}
\begin{equation}
1 = \frac{f^{2} N(0)}{2\delta\epsilon} \int_{0}^{\delta\epsilon }d\xi~ 
\frac{1}{\sqrt{\xi^2 + \Delta^2(T)}}
\tanh\left[ \frac{\sqrt{\xi^2 + \Delta^2(T)}}{k_B\,T} \right] \label{Eq5050-BCS}
\end{equation}
\end{small} 
with $\xi \equiv \epsilon_k - E_F$ the single energy particle relative to $E_F$. The 50-50 proportion implies that $E_f = E_F$, thus one can identify $\delta\epsilon = \hbar \omega_D$ with the Debye energy of the ionic lattice, and $f^2 N(0)/2\delta\epsilon \equiv \lambda_{BCS}$ the BCS dimensionless interaction parameter with $V = f^2/2\delta\epsilon$. In units of the Fermi energy, the BF interaction function can be related to a BF \textit{strength} interaction defined as $\tilde{G} \equiv f^{2} m^{3/2}/2^{5/2}\pi^{2}\hbar^{3}E_F^{1/2}$ and associate it to $\lambda_{BCS} = 2\tilde{G}/\delta\tilde{\epsilon}$.

Summarizing, if one solves \eqref{Eq5050} with DOS constant and with $\mu = E_F$ and a symmetric distribution between holes/electrons,  we recover the BCS energy gap equation (3.27) from Ref.~\cite{BCS}  and Eq.\eqref{BCS-lambda} for $T \to 0$, which is our first approximation within the ternary BF gas.

\begin{table*}[!htb]
\begin{center}
\caption{Shown are the four approximation cases for DOS and $\mu$ listed in the second and third columns, along with the calculated values of the critical temperature $T_c/T_F$. At zero temperature, the superconductor energy gap $\Delta(0)/E_F$, the difference of chemical potential $\Delta \mu(0)/E_F$, the difference of the SC thermodynamic potential with respect to its corresponding normal state $\Delta\Omega(0)/NE_F$ and the SC and normal Helmholtz free energy, i.e., $F_{\mathtt{s}}(0)/NE_F$ and $F_{\mathtt{n}}(0)/NE_F$. The exact normal free energy at $T = 0$ was taken as a reference for SC and normal free energy. Last column shows the condensation energies $E_{cond}(0)/NE_F$ with $\delta\tilde{\epsilon} = 10^{-3}$ and $\tilde{G} = 10^{-4}$ using \eqref{Eq-Econd}.}\label{Table-1}
\vspace*{0.15cm}
\resizebox{0.99\textwidth}{!}{\large 
\begin{tabular}{c|cc|ccccccc}
\toprule \toprule
Case &
DOS &
$\mu$ &
$T_c/T_F$  & 
$\Delta(0)/E_F$  & 
$\Delta \mu(0)/E_F$  &
$\Delta \Omega(0)/NE_F$  &
$F_{\mathtt{s}}(0)/NE_F$  &
$F_{\mathtt{n}}(0)/NE_F$  &
$E_{cond}(0)/NE_F$ \\
&
&
& $\times 10^{6}$ &
$\times 10^{5}$ &
$\times 10^{10}$ &
$\times 10^{10}$ &
$\times 10^{10}$ & 
$\times 10^{10}$ &
$\times 10^{10}$ \\
\midrule
i		& const & const  & 	\multicolumn{1}{l}{7.6399284}  & \multicolumn{1}{l}{1.34765058}	&	0  & \multicolumn{1}{l}{-0.6810446} & -3.1816074 & -2.5005627 & -0.6810447	\\

ii		& var   & const  & 	\multicolumn{1}{l}{7.639927977} & \multicolumn{1}{l}{1.347650498}	&	0  & \multicolumn{1}{l}{-0.6810441} & -0.6811381 & -0.0000940 & -0.6810441	\\

iii		& const & var    &	\multicolumn{1}{l}{7.6399282} & \multicolumn{1}{l}{1.34765054}	& -2500.6439366	 & \multicolumn{1}{l}{2499.9633635} & -3.1811367 & -2.5005635  & -0.6805731	\\

iv   	& var   & var    &	\multicolumn{1}{l}{7.639927976} & \multicolumn{1}{l}{1.347650497}	& -2.0434118	& \multicolumn{1}{l}{1.3623679} & -0.6811387 & -0.0000948 & -0.6810439	\\
\bottomrule \bottomrule
\end{tabular}
}
\end{center}
\end{table*}

\subsection{Other Approximations}

We now proceed to find the critical temperature and the superconducting energy gap for two other approximations (ii), (iii) and for the exact case (iv), which are needed to obtain the Helmholtz free energy and the condensation energy.

In the ternary BF theory the superconductor energy gap and chemical potential can be found by solving simultaneously  Eqs.\eqref{Eq5050} and \eqref{number5050}, taking the 50--50 proportions for each one of the approximations listed above, together with the typical values of conventional superconductors, namely, a generic case: Fermi temperature $T_F = 1.6\times 10^5 $K, Debye energy $\delta\tilde{\epsilon} = 10^{-3}$, where tilde means made dimensionless with Fermi energy, and the BF strength interaction $\tilde{G} = 10^{-4}$ which is related to $\lambda_{BCS} = 1/5$, resulting in the numerical values of the energy gap $\Delta(T)/E_F$; the critical temperature $T_c/T_F$; the  chemical potential difference $\Delta\mu(T)/E_F = [\mu_{\mathtt{s}}(T) - \mu_{\mathtt{n}}(T)]/E_F$, where $\mu_{\mathtt{n}}(T) = \mu_{IFG}(T)$ is the chemical potential of the ideal Fermi gas (IFG).  Thus, we substitute the above values  in \eqref{omega} for the thermodynamic potential difference $\Delta \Omega(T)/NE_F = [\Omega_{\mathtt{s}}(T) - \Omega_{\mathtt{n}}(T)]/NE_F$; in \eqref{Eq-Helmholtz} for the SC Helmholtz free energy $F_{\mathtt{s}}(T)/NE_F$ and in \eqref{Eq-Econd} for the condensation energy $E_{cond}(T)/NE_F= [F_{\mathtt{s}}(T) - F_{\mathtt{n}}(T)]/NE_F$. All of these quantities are evaluated at $T=0$ and shown in Table\,\ref{Table-1} for the four approximations analyzed.

\begin{figure}[!ht]
\centering
\includegraphics[width=7.5cm]{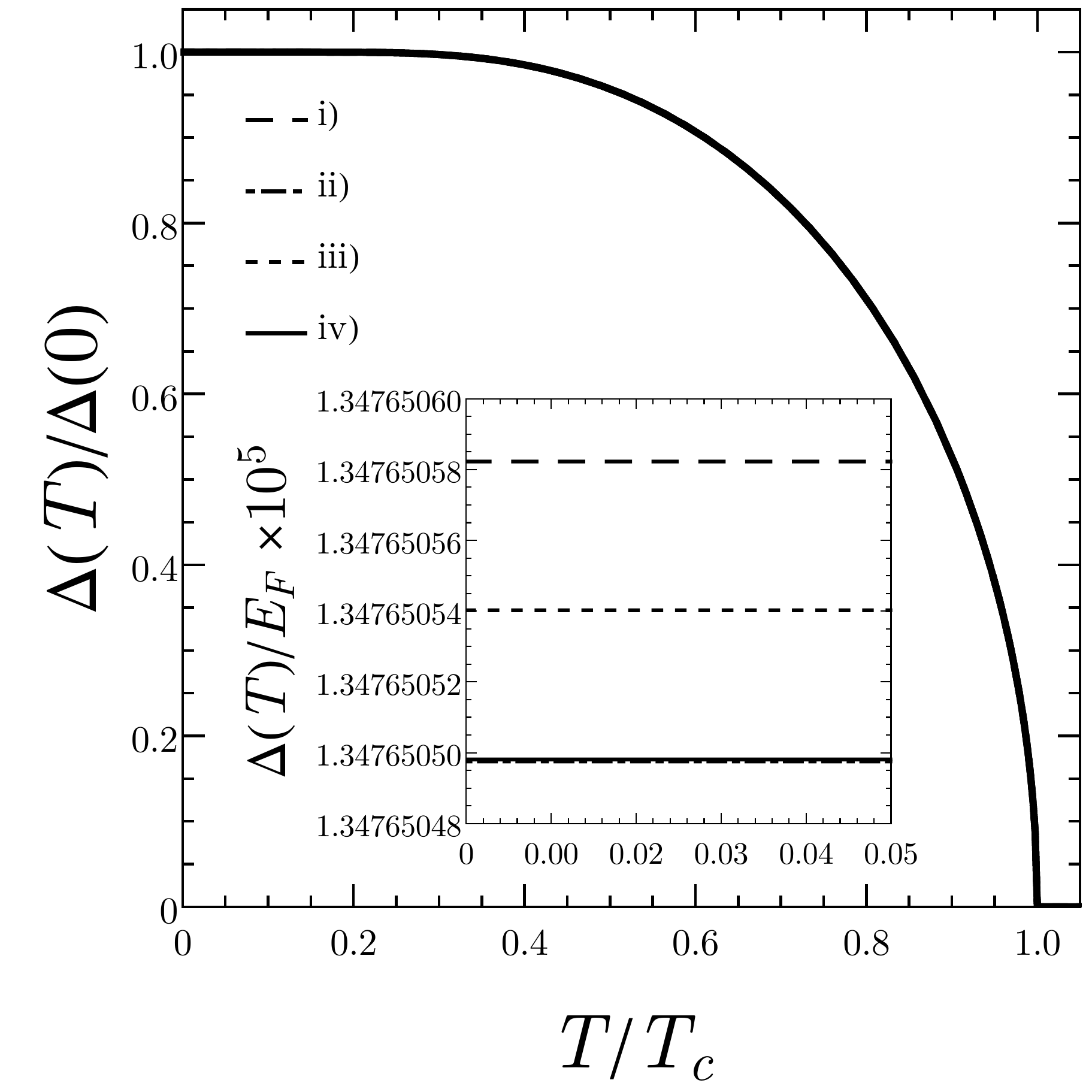}
\caption{Energy gap $\Delta(T)/\Delta(0)$ vs. $T/T_c$ using several approximations cited in text. Note that the behavior of all cases have the well-known half-bell shape, as expected. Inset shows a zoom of $\Delta(T)/E_F$ vs. $T/T_c$ at very low temperatures. Note that the difference between  cases (ii) and (iv) cannot be seen in the scale used. Here were used the 50--50 proportions and $\delta\tilde{\epsilon} = 10^{-3}$ and $\tilde{G} = 10^{-4}$ with $T_F = 1.6\times 10^{5}$ K.}\label{Fig-Gap}
\end{figure}

The curves of the energy gap as functions of temperature for each of the approximations listed above is plotted in Fig.~\ref{Fig-Gap}. As can be seen, the general behavior of the energy gap  between $0 \leq T \leq T_c$ is essentially the same, since there is no change in the half-bell shape of the curve. However, the numerical values at $T=0$, which are listed in Table~\ref{Table-1} and plotted in the Inset of Fig.~\ref{Fig-Gap}, show a tiny difference among the calculations, which occurs in the ninth digit of the difference between (i), (ii) and (iii), (iv), namely without or with chemical potential as constant.
\begin{figure}[!ht]
\centering
\includegraphics[width=7.5cm]{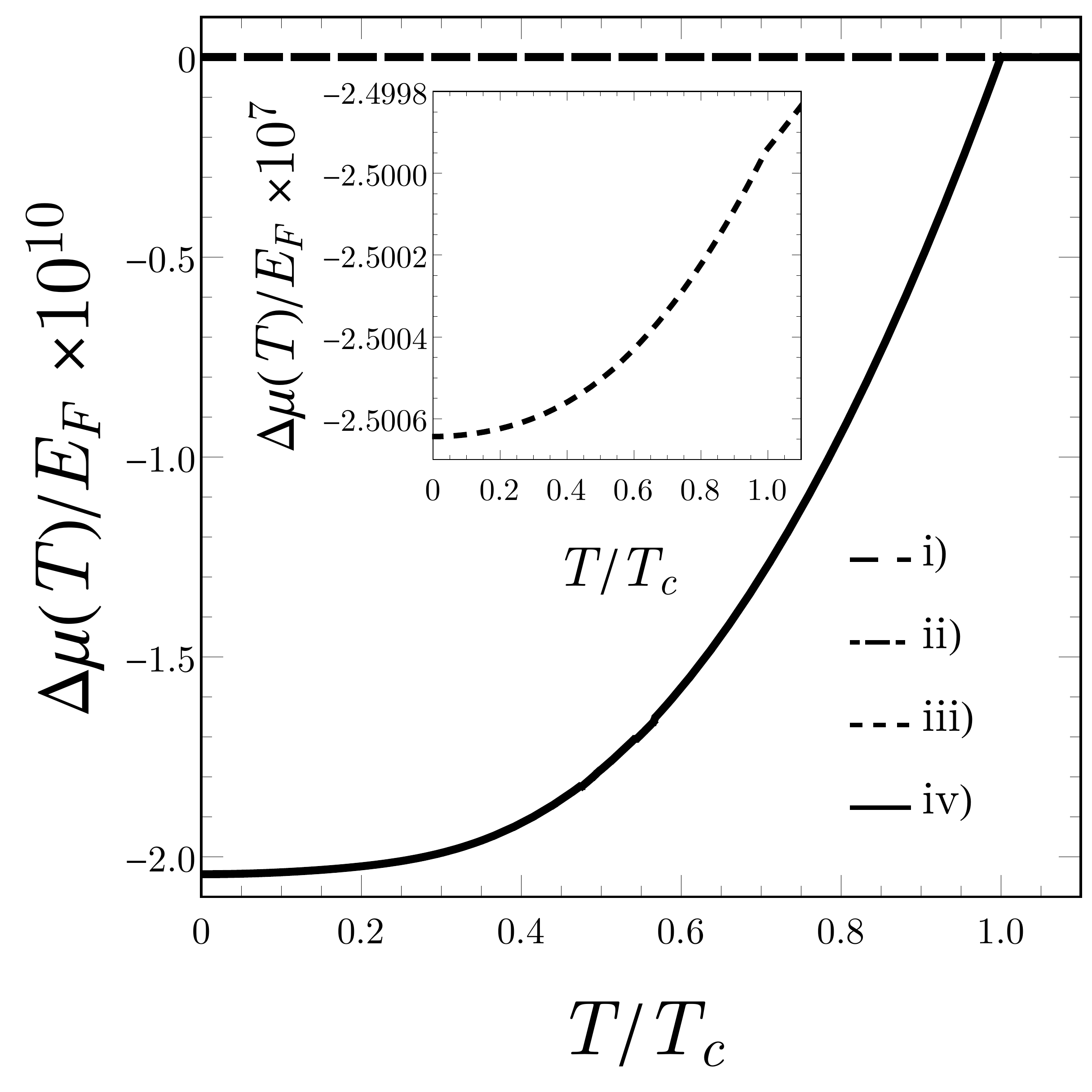}
\vspace*{-0.25cm}
\caption{$\Delta \mu(T)/E_F = [\mu_{\mathtt{s}}(T) - \mu_{\mathtt{n}}(T)]/E_F$ the difference of chemical potential with respect that of IFG vs. $T/T_{c}$ using the approximations BCS, (ii) and (iii) and the exact case as depicted in text. Inset shows $\Delta \mu(T)/E_F = [\mu_{\mathtt{s}}(T) - \mu_{\mathtt{n}}(T)]/E_F$ using the approximation (ii). When DOS is constant the difference of chemical potential increase at least 3 orders of magnitude with respect to the one with DOS variable. Note that the difference between  cases (i) and (ii) cannot be seen in the scale used. Here were used the 50--50 proportions and $\delta\tilde{\epsilon} = 10^{-3}$ and $\tilde{G} = 10^{-4}$ solving \eqref{Eq5050} with \eqref{number5050}.}\label{Fig-Mu}
\end{figure}

\begin{figure*}[!htb]
\centering
\subfigure[]{\includegraphics[width=7.5cm]{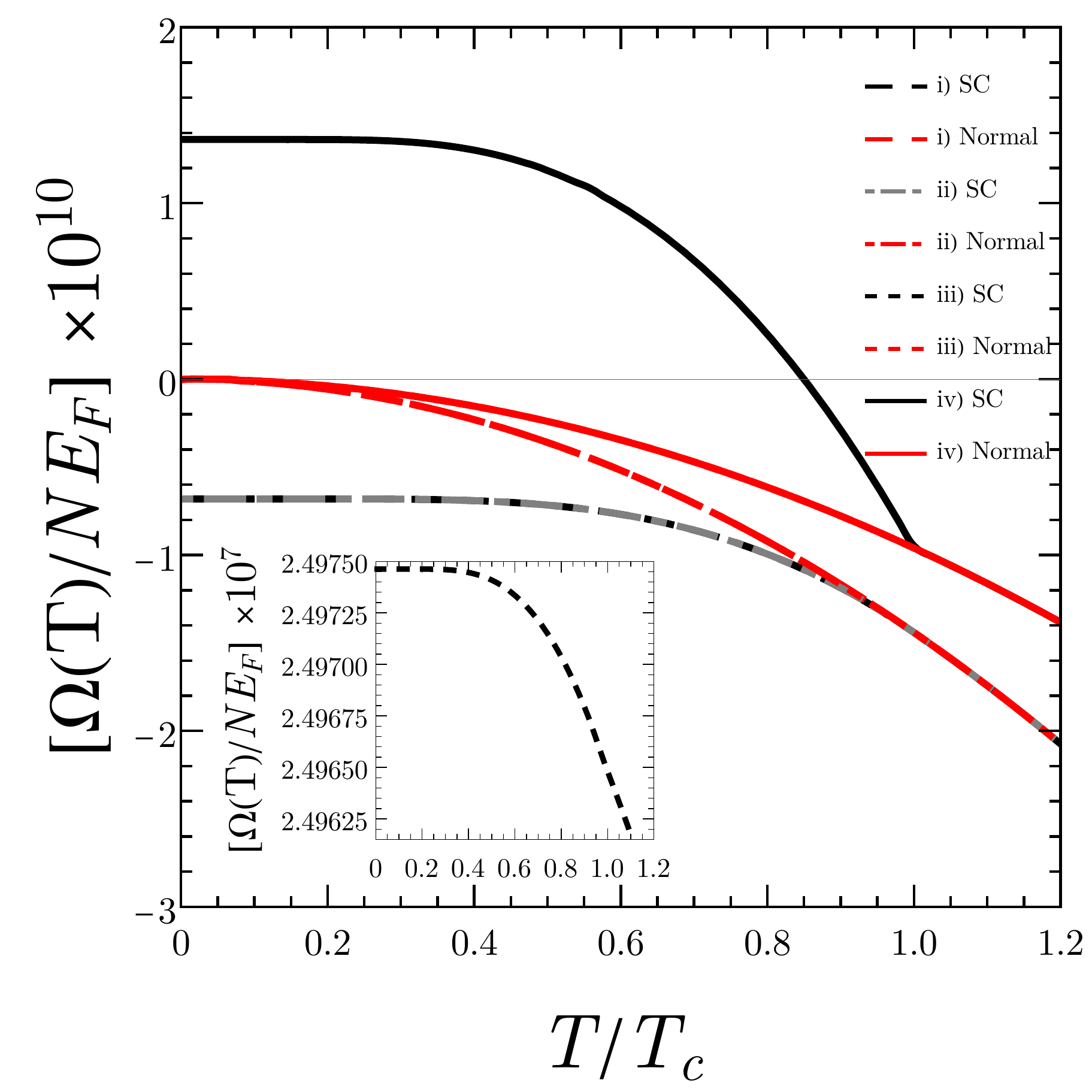}}
\qquad
\subfigure[]{\includegraphics[width=7.5cm]{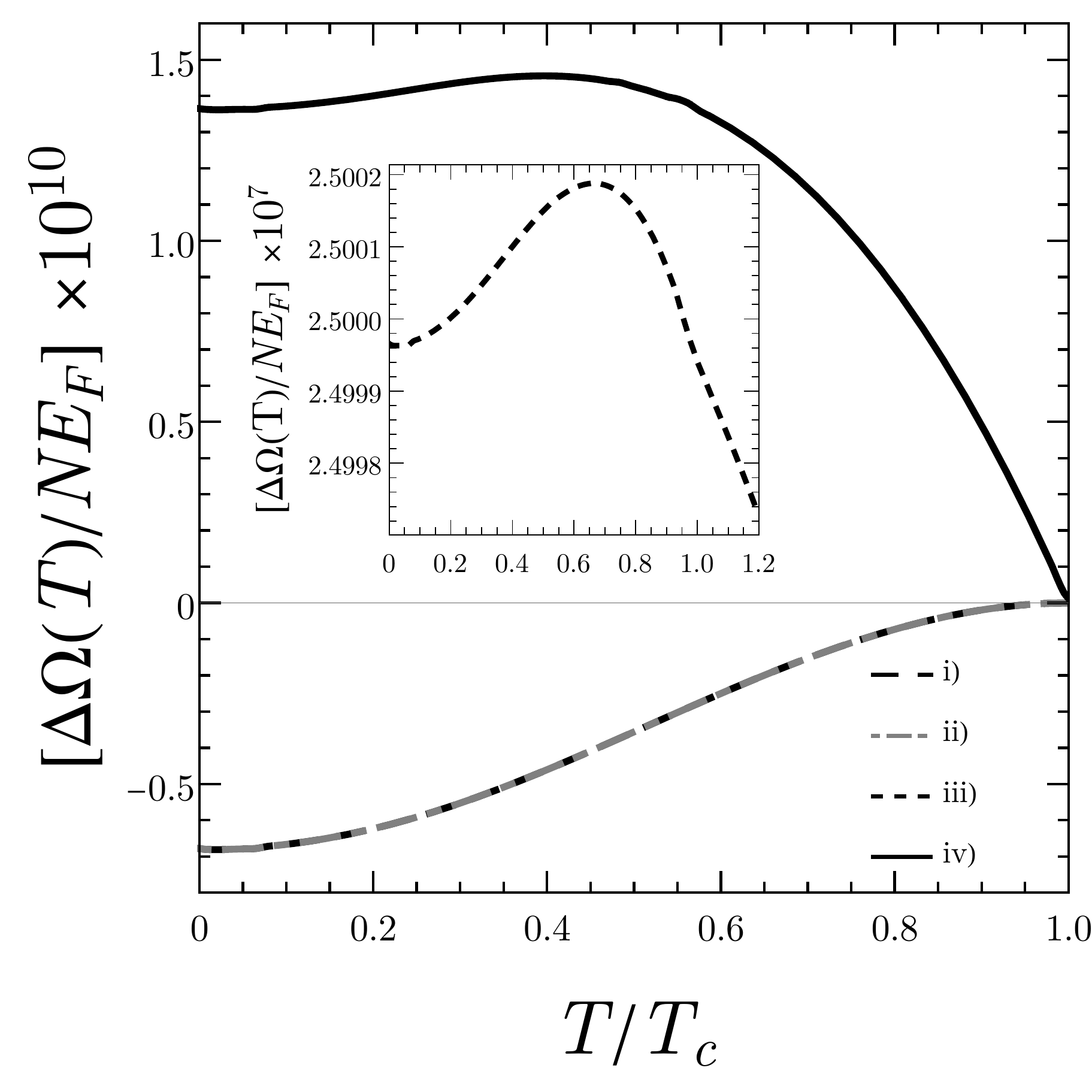}}
\caption{(Color online) a) Superconductor thermodynamic potential $\Omega_{\mathtt{s}}(T)/NE_F$  vs. $T/T_c$ along the normal thermodynamic potential $\Omega_{\mathtt{n}}(T)/NE_F$ (red curves) described in text. Cases (i), (ii) and (iv)  are plotted by black curves, while case (iii) by a gray curve. Here $\Omega_{\mathtt{s}}(T)/NE_F$ and $\Omega_{\mathtt{n}}(T)/NE_F$ were referenced to their corresponding $\Omega_{\mathtt{n}}(T=0)/NE_F=0$ for each case. Note that the (i) and (ii) cases have a negative sign while (iv) is positive. Inset shows the (iii) case, which is four orders of magnitude greater than the (iv) case.
b) Difference of thermodynamic potential $\Delta\Omega(T)/NE_F = [\Omega_{\mathtt{s}}(T) - \Omega_{\mathtt{n}}(T)]/NE_F$ for the same approximations. The black-dashed curve is the case (i), the gray-dashed curve is (ii), the short-dashed-black curve is (iii) and the full-black curve is (iv). Inset shows the (iii) case, where the normal thermodynamic potential has been enhanced five orders of magnitude with respect the SC thermodynamic potential. Note that the difference between  cases (i) and (ii) cannot be seen in the scale used.
}\label{Fig-Omega}
\end{figure*}

The behavior of the chemical potential as a function of temperature with the approximations enumerated above is shown in Fig.~\ref{Fig-Mu} by solving \eqref{Eq5050} and \eqref{number5050}. The plotted curves are the difference between superconductor and IFG chemical potentials, i.e., $\Delta \mu(T)/E_F = [\mu_{\mathtt{s}}(T) - \mu_{\mathtt{n}}(T)]/E_F$. At $T=0$ there is a very tiny difference between superconductor and normal chemical potentials, but very important when we calculate the Helmholtz free energy and the condensation energy.

\section{Helmholtz free energy and condensation energy}

\begin{figure*}[!ht]
\centering
\subfigure[]{\includegraphics[width=7.5cm]{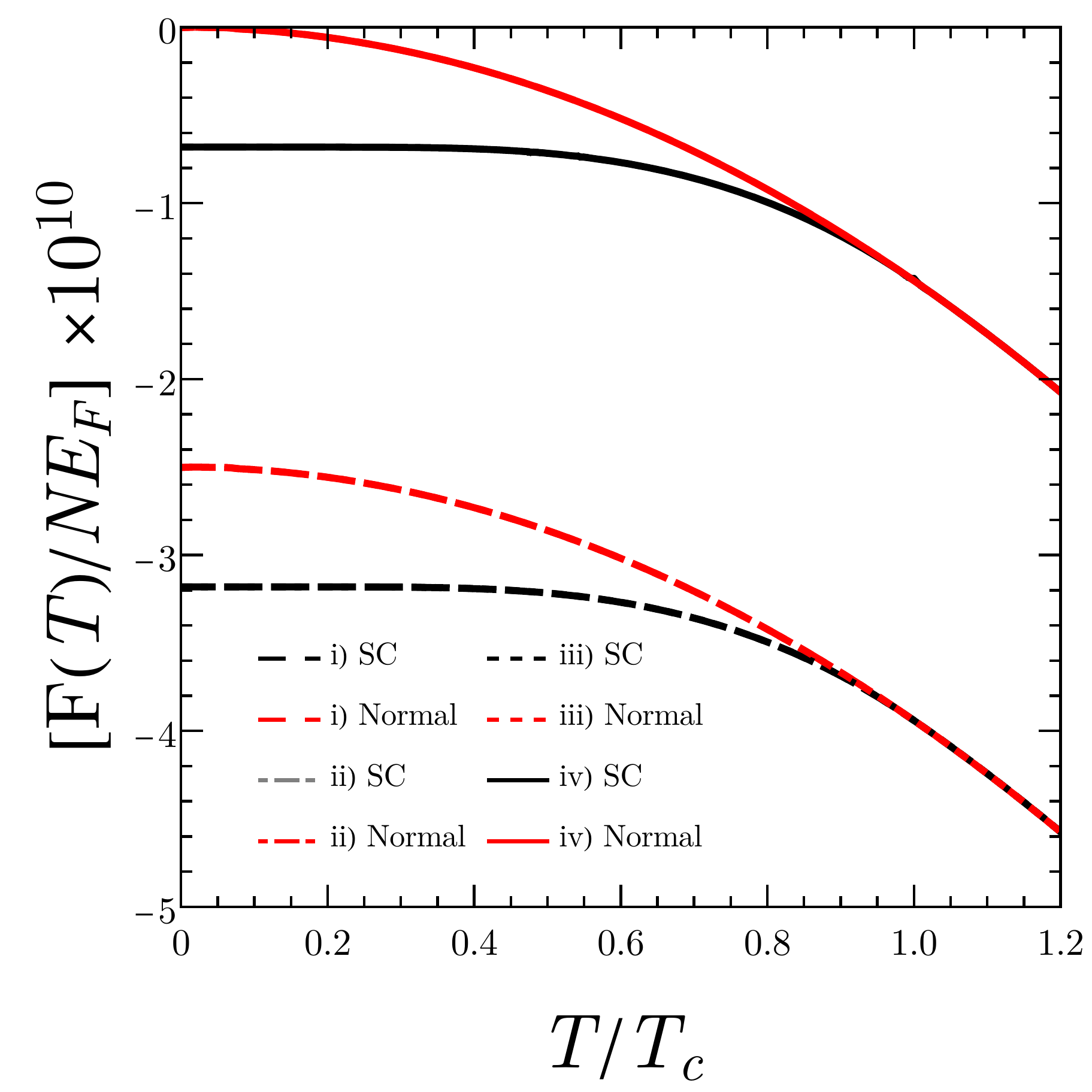}}
\qquad
\subfigure[]{\includegraphics[width=7.5cm]{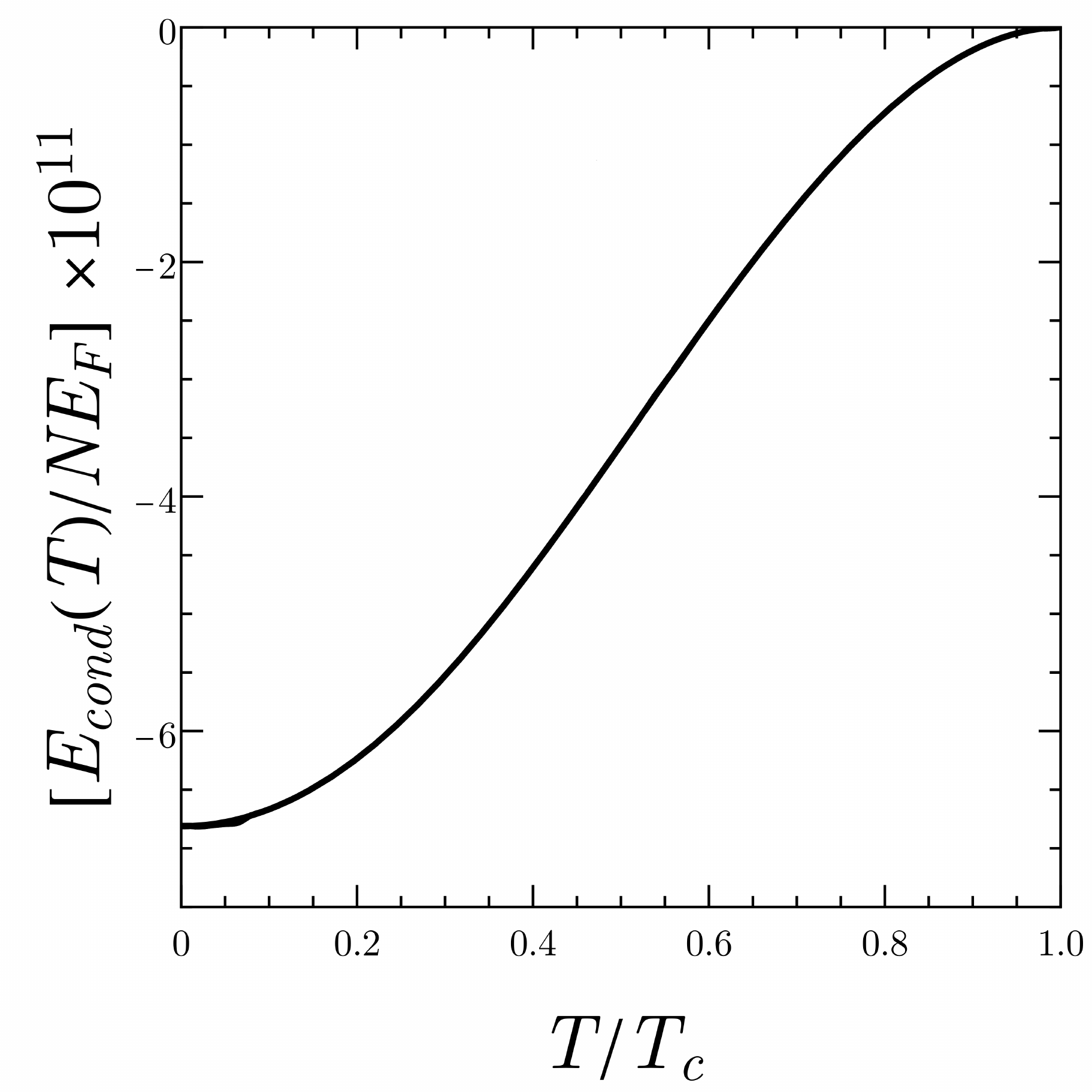}}
\caption{(Color online) \textbf{a}) Helmholtz free energy $F(T)/NE_F$ vs. $T/T_c$ for SC and normal states, black and red curves, respectively, for each approximation as described in the text.
Note that cases (i) and (iii)  overlap, although they differ at $T=0$ in the fourth figure of calculation, according to Table~\ref{Table-1}, while cases (ii) and (iv) also overlap  but differ in the seventh figure at $T=0$.
\textbf{b}) Condensation energy $E_{cond}(T)/NE_F$ vs. $T/T_c$ for all cases. All the cases lie very close to each other, as can be seen from Table~\ref{Table-1}.}\label{Fig-Helmholtz-Econd}
\end{figure*}
Fig.~\ref{Fig-Omega}a shows the grand thermodynamic potential $\Omega(T)/NE_F$ for the normal and SC states, red and black curves respectively, and in Fig.~\ref{Fig-Omega}b we plot the difference of the thermodynamic potential $\Delta\Omega(T)/NE_F$ for the three approximations just mentioned plus the exact case, by using the energy gap and chemical potential values previously calculated. The corresponding normal state at $T=0$ has been chosen for each case as the zero of coordinates. At first glance,  the normal-state approximations (i), (ii) and (iii) are different from the exact case (iv). When the chemical potential is taken as constant, $\Omega_{\mathtt{s}}(T)/NE_F < 0$ for the (i) and (ii) SC states, for all temperatures, while the behavior changes when the chemical potential is variable, cases (iii) and (iv), leading to $\Omega_{\mathtt{s}}(T)/NE_F >0$, as  illustrated in Fig.~\ref{Fig-Omega}b. Thus, the difference of thermodynamic potentials is positive for the (iii) and (iv) cases and negative for the (i) and (ii) cases; this difference coincides for the last two cases.

On the one hand, if $\mu$ is constant, $\Delta\Omega(T)/NE_F$ is negative with the same order or magnitude of the condensation energy for the (i) and (ii) cases, as pointed out in Ref.~\cite{fetter} and reported here in Table~\ref{Table-1}; on the other hand, when $\mu$ is taken as variable, cases (iii) and (iv) of $\Delta\Omega(T)/NE_F$ is positive and greater than the cases (i) and (ii); however it should be taken like this in order to calculate the correct value of the condensation energy.

In Fig.~\ref{Fig-Helmholtz-Econd}a the Helmholtz free energy $F(T)/NE_F$ for each case is plotted,
red curves are labeled for normal state while black curves for SC state. One can see that all cases follow the expected behavior, although there is a huge difference between  cases (i) and (iii), as well as between (ii) and (iv), both  in the thermodynamic potential and the chemical potential differences. In Fig.~\ref{Fig-Helmholtz-Econd}b the generic condensation energy is shown for all cases, with $\delta\tilde{\epsilon} = 10^{-3}$ and $\tilde{G} = 10^{-4}$. All cases describe the general behavior of  the condensation energy, although as one can see in Table~\ref{Table-1}, results of the last column at $T=0$ differ in the sixth figure of the calculation. This suggest that the condensation energy curve also changes slightly in $0 \leq T \leq T_c$.

\begin{figure}[!ht]
	\centering
	\includegraphics[width=7.5cm]{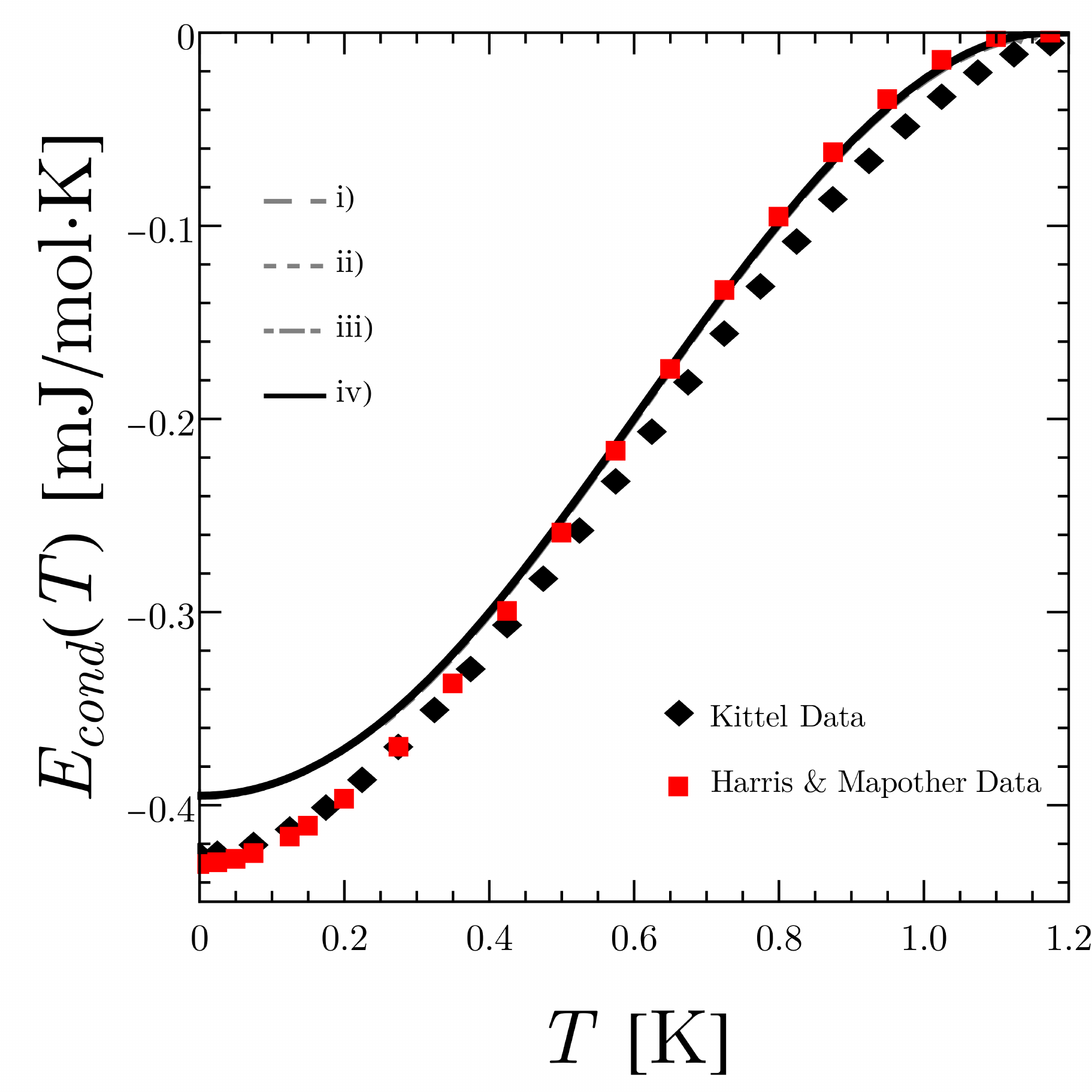}
	\caption{Condensation energy vs. $T$ [K] for aluminum superconductor compared with approximation cases listed in the text. All cases follow data trends, which were calculated here from \cite{kittel, harris} as described in the text. For (i) and (ii) curves the BCS energy gap solution at $T=0$ was used, namely $\Delta(0)/E_F = 1.5 \times 10^{-5}$, while for (iii) and (iv) cases $\Delta(0)/E_F = 1.48 \times 10^{-5}$ was used.
Here were used $T_{F} = 1.36 \times 10^5$ K \cite{poole}, $\delta\tilde{\epsilon} = 3\times 10^{-3}$ \cite{poole} and $\tilde{G} = 2.5\times 10^{-4}$, thus obtaining a critical temperature of $T_c = 1.17$ K.
}\label{Fig-EcondAl}
\end{figure}

The general behavior of the energy gap of aluminum is depicted by all cases and follows the data trends; these cases correctly predict the critical temperature $T_c = 1.17$ K \cite{roberts}, although the energy gap lies in the range $\Delta(0) = [0.1784418, 0.1784421]$ meV while the experimental data \cite{biondi} is $\Delta(0) = 0.163$ meV. In Fig.~\ref{Fig-EcondAl} we show the above approximations and compare with the condensation energy of aluminum superconductor data.  Data from SC and normal Helmholtz free energy curves for Al was taken from Ref.~\cite{kittel}, and interpolated (adding both curves) to obtain the condensation energy curve reported here. While data for the critical field was taken  from Ref.~\cite{harris} by using the thermodynamic relation \cite{shoenberg} $\Delta S = S_n - S_s = -(V H_c/4\pi) (\partial H_c)/(\partial T)_P$ and $E_{cond}(T) = \int_{0}^{T_c} dT \Delta S$ to obtain the condensation energy.

This is so by several aspects of this BF theory, e.g., it has been taken with only the interactions between e-e and h-h to form 2e/2hCPs, so the following interactions  have been ignored in this work: e-h, e-2eCP, h-2hCP, 2eCP-2hCP. Furthermore, the BF energy-gap solution at $T=0$ is slightly different from experimental data, meaning that the energy gap at $T=0$ leads to find out the condensation energy. One can then see that the condensation energy values at $T=0$, are $E_{cond}(0) = -0.385$ [mJ/mol·K], which lie slightly above the experimental data $E_{cond}(0) = -0.42$ [mJ/mol·K].

\section{Conclusions}

We highlighted the effect of taking DOS and $\mu$ either constant or variable on the condensation energy. To do this, we used a ternary BF superconducting theory, composed of unbound electrons/holes as fermions with two-electron and two-hole Cooper pairs as bosons. We are able to calculate the energy gap, the chemical potential, the thermodynamic potential, the Helmholtz free energy and the condensation energy with 50--50 proportions between 2e/2hCPs. Essentially the energy gap has the same behavior and order of magnitude as the ones obtained taking different approximations, but the difference of thermodynamic potential behaves very differently.

The first approximation (DOS and $\mu$ constant) taken by BCS theory cannot predict the thermodynamic potential, which is basic for calculating other thermodynamic properties, as the entropy or the specific heat. While the (iv) approximation, i.e., the exact case taken here with the ternary BF gas mixture leads to the correct way to calculate the free and condensation energy. One can see that the BCS theory results are recovered by this ternary BF theory with a 50--50 proportions.

To compare our theoretical results we used the aluminum superconductor condensation energy reported experimental data \cite{kittel, harris}, the general behavior is quite well in following data trends. An important result here is that when one takes DOS and $\mu$ as constants, the difference of the thermodynamic potential is equal to the condensation energy for all temperatures, although the correct calculation comes from taking DOS and $\mu$ variable, leading to the right free-energy value, an also to the correct condensation energy. All cases describe well the general behavior of the condensation energy of aluminum, but it is slightly different at $T=0$, this is so among other things, due to the leading term in the free energy. It is worth mentioning here that a DOS constant must be taken in both, energy gap and number equations to obtain such values, otherwise there is no critical temperature and energy gap, or there is no superconductor state. We conclude that taking DOS and the chemical potential as variables, the condensation energy values match  the experimental data reported.

\section*{Acknowledgments}
IC thanks CONACyT for Postdoc grant EPA1 \#\,869450. PS and MAS thank PAPIIT-DGAPA-UNAM for grant IN110319. MdeLl thanks to PAPIIT-DGAPA-UNAM for grant IN115120. We thank O.A. Rodr\'{i}guez for his computer support in the development of this paper.

\appendix
\section{BCS Helmholtz free and condensation energies}

We recall here some of the BCS equations to compare with the corresponding BF free energy and condensation energy.
The SC Helmholtz free energy of the BCS \cite{BCS} theory per unit volume is
\begin{flalign}
F_{\mathtt{s}} &= -2N(0) \int\limits_{0}^{\infty} d\epsilon_k\, \left( \frac{2 \epsilon_k^{2} + \Delta^{2}}{\sqrt{\epsilon_k^{2} + \Delta^{2}}} \right)
\left(\exp \{ \frac{\sqrt{\epsilon_k^{2} + \Delta^{2}}}{k_B\,T}\} +1  \right)^{-1}
\notag \\
&-
N(0) (\hbar \omega_D)^{2} \left\lbrace \sqrt{1 + (\Delta/\hbar\omega_D)^2} -1 \right\rbrace \notag
\end{flalign}
and for the normal state is
\begin{eqnarray}
F_{\mathtt{n}} &=& -4N(0) k_B\,T \int\limits_{0}^{\infty} d\epsilon \ln [1+ \exp(-\beta \epsilon)]
\notag \\
&=& - \tfrac{1}{3} \pi^{2}N(0) (k_B\,T)^2.
\end{eqnarray}
Therefore, the condensation energy (\ref{Eq-Econd}) is

\begin{flalign}
E_{cond} &= F_{\mathtt{s}}-F_{\mathtt{n}} \notag \\
&= -N(0) (\hbar \omega_D)^2 
\left\{\left[\sqrt{\left(\frac{\Delta(T)}{\hbar \omega _D}\right)^2+1}\right]-1\right\}
\notag \\
&+ \frac{\pi ^2}{3} N(0) \left(k_B\, T\right)^2 
\bigg[
1-\left(\frac{1}{k_B\,T}\right)^2 \int\limits_{0}^{\infty } d\epsilon_k 
\left( \frac{2 \epsilon_k ^2+ \Delta(T)^2}{\sqrt{\epsilon_k ^2+ \Delta(T)^2}}
\right) 
\notag \\
&\times
\left(\frac{1}{\exp\{\frac{\sqrt{\epsilon_k ^2+ \Delta(T)^2}}{k_B\,T}\}+1}\right) 
\bigg]. \label{EcondBCS}
\end{flalign}

\noindent The energy gap values obtained from solving Eq. (3.27) of Ref.\,\cite{BCS} must be introduced in \eqref{EcondBCS}, by taking the approximation (i) with DOS constant and $\mu = E_F$.

In some textbooks, like Fetter and Walecka \cite{fetter}, it is common to find the condensation energy as the difference between the thermodynamic potentials,  as in their Eq. (51.53), ignoring the contribution coming from the chemical potential difference, as
%
\begin{small}
\begin{flalign}\label{EcondFW}
\frac{\Omega_{\mathtt{s}} - \Omega_{\mathtt{n}}}{L^3} &= -\tfrac{1}{2}N(0) \Delta^2 - N(0) \Delta^{2}\ln \left( \frac{\Delta_0}{\Delta} \right)
\\
&- 4 N(0) k_B\,T \int\limits_{0}^{\hbar\omega_D} d\xi \ln (1+ \exp\{\tfrac{E}{k_B\,T} \})
+ \tfrac{1}{3} \pi^{2}N(0) (k_B\,T)^{2} \notag
\end{flalign}
\end{small}
%
where the energy gap values must be substituted in order to solve this expression. Fetter and Walecka reinterpret \cite{fetter} the above expression as their Eq. (51.54), namely 
\begin{eqnarray}
\frac{F_{\mathtt{s}} - F_{\mathtt{n}}}{L^3} \underset{T \to 0}{\approx} -\tfrac{1}{2}N(0) \Delta_{0}^2 + \tfrac{1}{3} \pi^{2}N(0) (k_B\,T)^{2},  \label{EcondFWT0}
\end{eqnarray}
which is the Helmholtz free energy when $T \to 0$. But their Eq. (51.63) shows
\begin{eqnarray}
\frac{\Omega_{\mathtt{s}} - \Omega_{\mathtt{n}}}{L^3} = -\frac{8}{7 \zeta(3)} N(0) (\pi\,k_B\,T_c)^{2}\,\frac{1}{2}\left(1- \frac{T}{T_c} \right)^{2} \label{EcondFWTc}
\end{eqnarray}
this is an explicit expression near $T_c$. The remarkable thing of Eqs. (51.53) and (51.63) is that the chemical potential is missing. They assume that the chemical potential difference contributes by only a very small amount $(\mu_{\mathtt{s}} - \mu_{\mathtt{n}})/\mu_{\mathtt{n}} = \delta$. Assuming \cite[p.335]{fetter} a single-particle spectrum $\epsilon_k \sim \Delta_0$, thus
\begin{equation}
\delta = -\frac{1}{3} \left( \frac{\Delta}{\epsilon_F}\right)^{2} \left[ 1 + 6\left( \frac{\partial \ln \Delta}{\partial \ln N} \right) \right]
\end{equation}
which justifies the omission of the $\delta^{2}$ correction in Eq. (37.52), namely
\begin{eqnarray}
E_{\mathtt{s}} - E_{\mathtt{n}}  &=& U_{\mathtt{s}}(\mu_{\mathtt{s}}) - U_{\mathtt{n}}(\mu_{\mathtt{n}}) + N \left(\mu_{\mathtt{s}} - \mu_{\mathtt{n}} \right)
\notag \\
&=& U_{\mathtt{s}}(\mu_{\mathtt{s}}) - U_{\mathtt{n}}(\mu_{\mathtt{n}}) 
\notag \\
&+& \delta \mu_n \left[\left(\frac{\partial U_s}{\partial \mu} \right)_{\mu_n} + N \right] + O(\delta^{2})
\notag \\
&\approx & U_{\mathtt{s}}(\mu_{\mathtt{s}}) - U_{\mathtt{n}}(\mu_{\mathtt{n}}).
\end{eqnarray}
However
\begin{equation}
\frac{N (\mu_{\mathtt{s}} - \mu_{\mathtt{n}})}{E_{\mathtt{s}} - E_{\mathtt{n}}} = \frac{1}{3} \left[ 1 + 6\left( \frac{\partial \ln \Delta}{\partial \ln N} \right) \right]
\end{equation}
is comparable with one; this implies
\begin{equation}
\delta \simeq -\left( \frac{\Delta}{E_F} \right)^{2}
\end{equation}
if the energy gap is $\Delta \sim 10^{-5} E_F$ thus $\delta \approx -10^{-10}$ at $T=0$. This agrees with van der Marel \cite{vandermarel} and more recently with Chávez \textit{et al}. \cite{chavez22} that the difference between the SC chemical potential and the normal chemical potential is finite and of the order of magnitude of $\delta$.

\section{Thermodynamic properties for aluminum}
\counterwithin{figure}{section}
\counterwithin{table}{section}
\setcounter{figure}{0}
\setcounter{table}{0}

In this section we show the thermodynamic properties for aluminum superconductor using the ternary BF model as described in section III, where we showed the same thermodynamic properties for the generic case. Energy gap and chemical potential values were obtained by solving simultaneously \eqref{Eq5050} and \eqref{number5050} with the dimensionless Debye energy of aluminum $\delta\tilde{\epsilon} = 3\times 10^{-3}$, the BF strength interaction $\tilde{G} = 2.5\times 10^{-4}$ and $T_F = 1.36 \times 10^{4}$ K.

Table~\ref{AlTable-1} shows the critical temperature $T_c/T_F$, at $T=0$ the energy gap $\Delta(0)/E_F$, chemical potential $\mu(0)/E_F$, thermodynamic potential $\Omega(0)/NE_F$, the SC and normal Helmholtz free energy $F_s(0)$, $F_n(0)/NE_F$, respectively as well as the condensation energy $E_{cond}(0)$.
\begin{table*}[!ht]
\begin{center}
\caption{Aluminum superconductor with the four approximation cases. DOS and $\mu$ listed in the first two columns, along with the calculated values of the critical temperature $T_c/T_F$, the superconductor energy gap $\Delta(0)/E_F$, the difference of chemical potential $\Delta \mu(0)/E_F$, the difference of the SC thermodynamic potential with respect to its corresponding normal state $\Delta\Omega(0)/NE_F$, the SC and normal Helmholtz free energy $F_{\mathtt{s}}(0)/NE_F$, $F_{\mathtt{n}}(0)/NE_F$ and the condensation energy $E_{cond}(0)/NE_F$ with $\delta\tilde{\epsilon} = 3\times 10^{-3}$ and $\tilde{G} = 2.5\times 10^{-4}$ by using \eqref{Eq-Econd}.}\label{AlTable-1}
\vspace*{0.25cm}
\resizebox{0.9\textwidth}{!}{\small
\begin{tabular}{ccccccccc}
\toprule \toprule
Case &
$T_c/T_F$  & 
$\Delta(0)/E_F$  & 
$\Delta \mu(0)/E_F$  &
$\Delta \Omega(0)/NE_F$  &
$F_{\mathtt{s}}(0)/NE_F$  &
$F_{\mathtt{n}}(0)/NE_F$  &
$E_{cond}(0)/NE_F$ &
$E_{cond}(0)$ \\ \\
& $\times 10^{6}$ &
$\times 10^{5}$ &
$\times 10^{10}$ &
$\times 10^{10}$ &
$\times 10^{10}$ & 
$\times 10^{10}$ &
$\times 10^{10}$ &
[mJ/mol·K] \\
\midrule
i		  & 	8.65 & \multicolumn{1}{l}{1.5232412}	  &	0  & \multicolumn{1}{r}{-0.86999}
& \multicolumn{1}{r}{-68.40810} & \multicolumn{1}{l}{-67.53811} & -0.86999 & -0.39549	\\

ii		  & 	8.65 & \multicolumn{1}{l}{1.523240419}	&	0  & \multicolumn{1}{r}{-0.87009}
& \multicolumn{1}{r}{-0.87009} & 0.00000 & -0.87009	& -0.39554 \\

iii		    &	8.65 & \multicolumn{1}{l}{1.5232408}	& -22511.55276	 & 22510.72081
& \multicolumn{1}{r}{-68.37014} & \multicolumn{1}{l}{-67.53819}  & -0.83195 &	-0.37820 \\

iv   	    &	8.65 & \multicolumn{1}{l}{1.523240418}	& \multicolumn{1}{r}{-3.17645}	& \multicolumn{1}{r}{2.30653}
& \multicolumn{1}{r}{-0.87010} & -0.00018 & -0.86992  & -0.39546	\\
\bottomrule \bottomrule
\end{tabular}
}
\end{center}
\end{table*}

\begin{figure}[H]
\centering
\includegraphics[width=7.5cm]{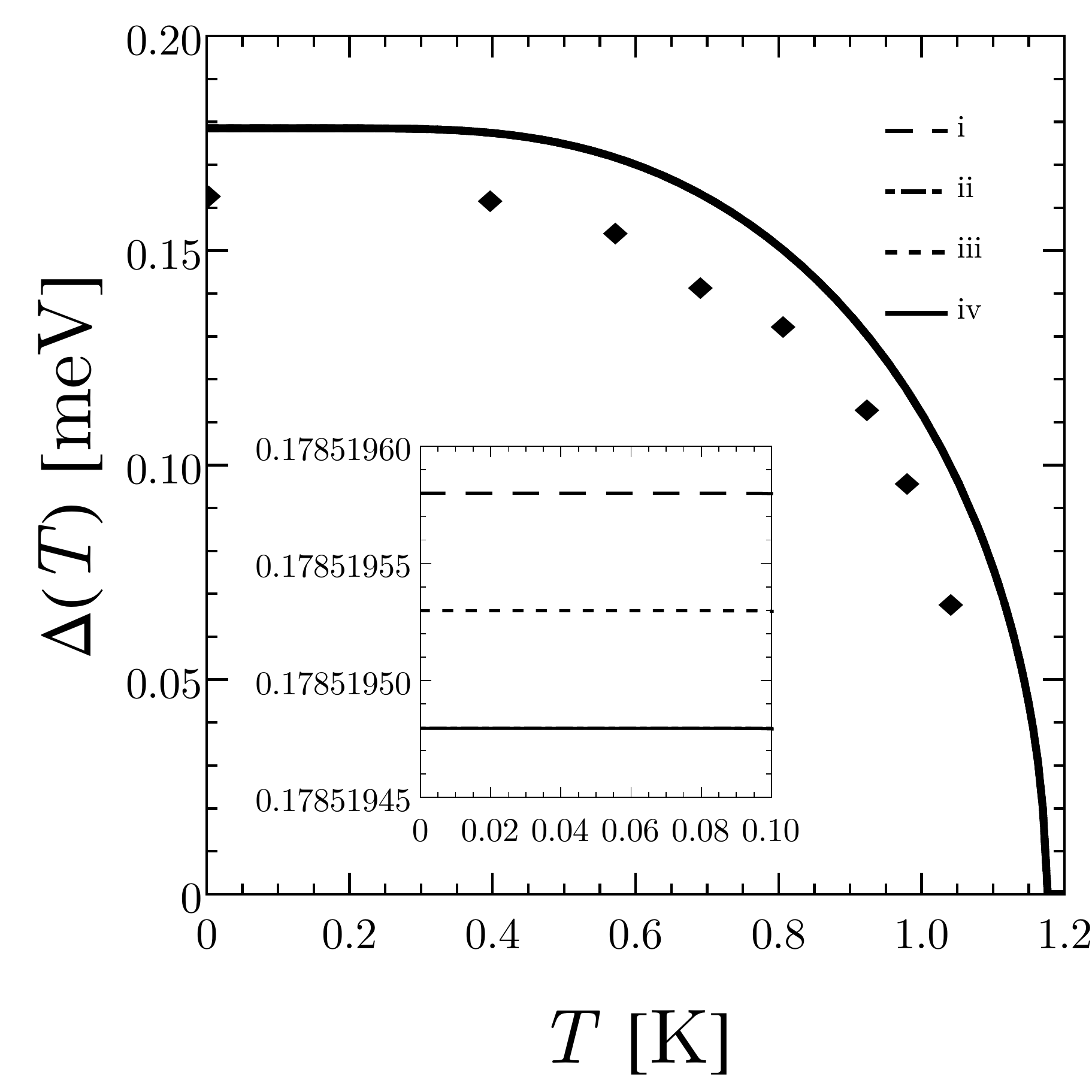}
\caption{Energy gap $\Delta(T)$ [meV] vs. $T$ [K] for aluminum superconductor compared with experimental data \cite{biondi}. We reproduce the critical temperature $T_c = 1.17$ K while the experimental energy gap is $\Delta(0) = 0.163$ [meV]. Inset shows the four approximations for low temperatures. Note that the energy gap values for (ii) and (iv) cases overlap and differ in the ninth figure of calculations as described in Table~\ref{AlTable-1}.}\label{Fig-GAP-Al}
\end{figure}

Fig.~\ref{Fig-GAP-Al} is plotted the energy gap $\Delta(T)$ [meV] vs. $T$ [K]. The four approximations follow the general behavior of the energy gap, although at $T=0$ it is slightly different from the reported data \cite{biondi} $\Delta(T) = 0.163$ [meV]. The energy gap and chemical potential values obtained above were substituted in \eqref{omega} in order to obtain the thermodynamic potential and in \eqref{Eq-Helmholtz} to obtain the Helmholtz free energy for aluminum.

\begin{figure*}[!ht]
\centering
\subfigure[]{\includegraphics[width=7.35cm]{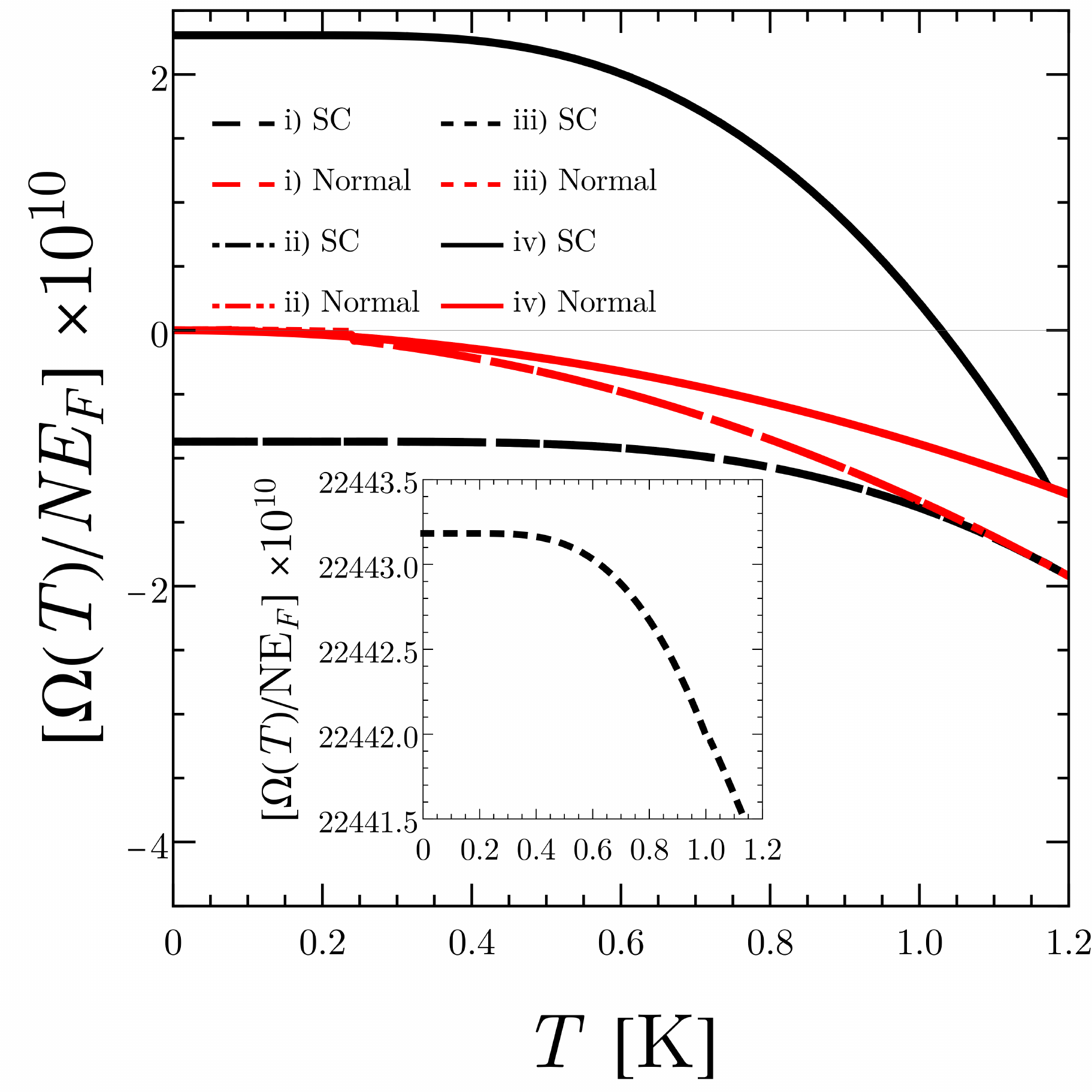}} \qquad
\subfigure[]{\includegraphics[width=8.45cm]{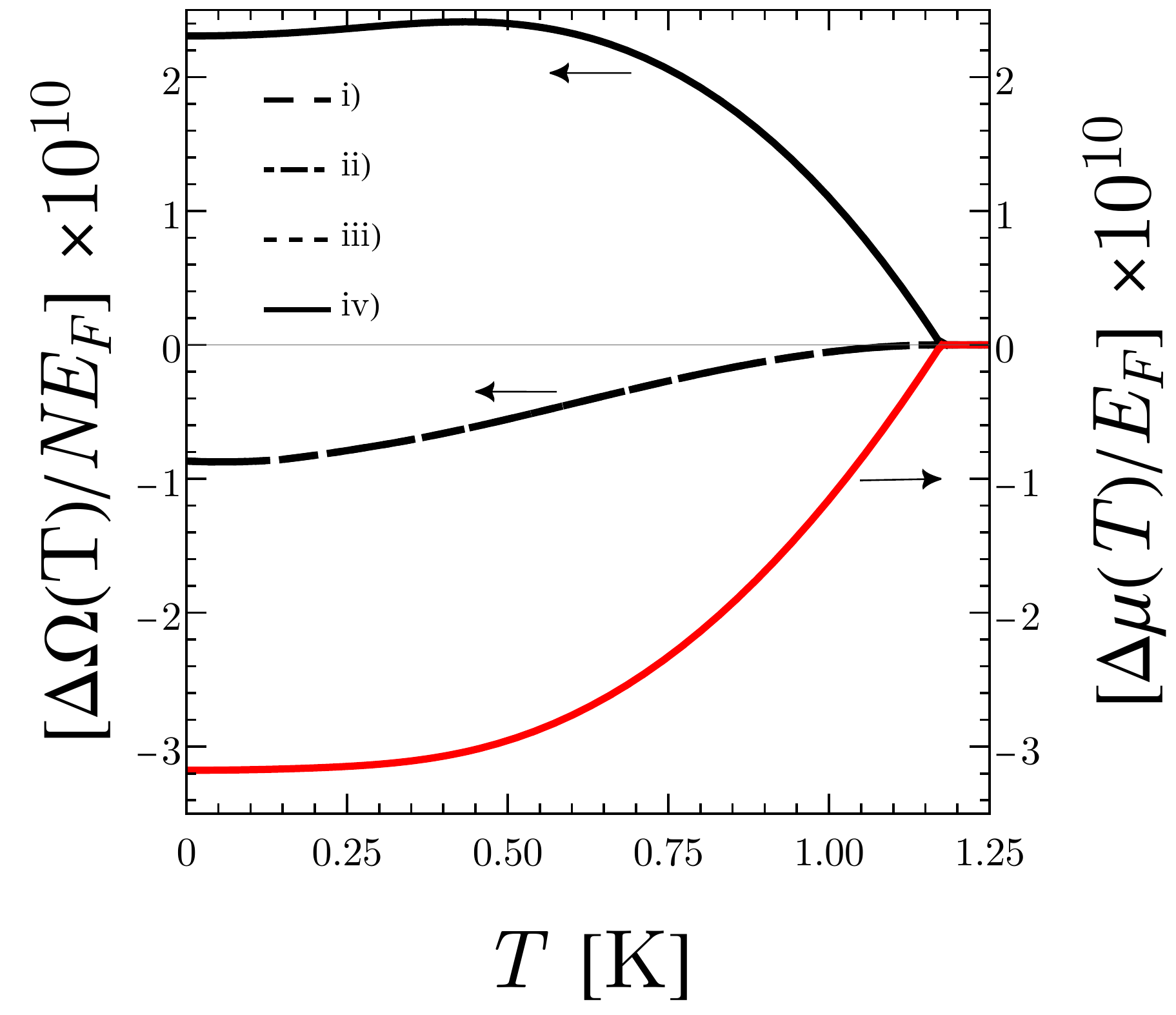}}
\caption{(Color online) (a) Thermodynamic potential $\Omega(T)/NE_F$ vs. $T$ [K] for aluminum with four cases as cited in text. The exact case (black curve) is above with respect to the normal state, while the other cases (black- and gray-dashed) curves are below from the normal state (red-dashed) curves. Inset shows the thermodynamic potential for the (iii) case. Note the five orders of magnitude with respect to the other cases.
(b) Difference of thermodynamic potential $\Delta\Omega(T)/NE_F = [\Omega_{\mathtt{s}}(T) - \Omega_{\mathtt{n}}(T)]/NE_F$ vs. $T$ [K]. For cases (i), (ii) and (iii) the difference is negative while for case (iv) it is positive. Also plotted is the difference of the chemical potential $\Delta\mu(T)/E_F = [\mu_{\mathtt{s}}(T)-\mu_{\mathtt{n}}(T)]/E_F)$ (red curve) for the exact case for comparison purposes. Here were used the energy gap and chemical potential values calculated above.}\label{Fig-Omega-Al}
\end{figure*}

Thus, in Fig.~\ref{Fig-Omega-Al}a is plotted the thermodynamic potential $\Omega(T)/NE_F$ vs. $T$ [K] and in Fig.~\ref{Fig-Omega-Al}b the difference of thermodynamic potential $\Delta\Omega(T)/NE_F = [\Omega_{\mathtt{s}}(T) - \Omega_{\mathtt{n}}(T)]/NE_F$ vs. $T$ [K], also plotted is the difference of the chemical potential $\Delta\mu(T)/E_F = [\mu_{\mathtt{s}}(T)-\mu_{\mathtt{n}}(T)]/E_F)$ for the exact case for comparison purposes only. As well as in the generic case, the thermodynamic potentials have been referenced to the normal state at $T=0$. From Table~\ref{AlTable-1}  one can see that the difference of thermodynamic and chemical potentials at $T=0$ has the same order of magnitude, leading to the correct calculation of the condensation energy. Note that both curves, have almost the same behavior with their corresponding sign.

\begin{figure}[H]
\centering
\includegraphics[width=7.5cm]{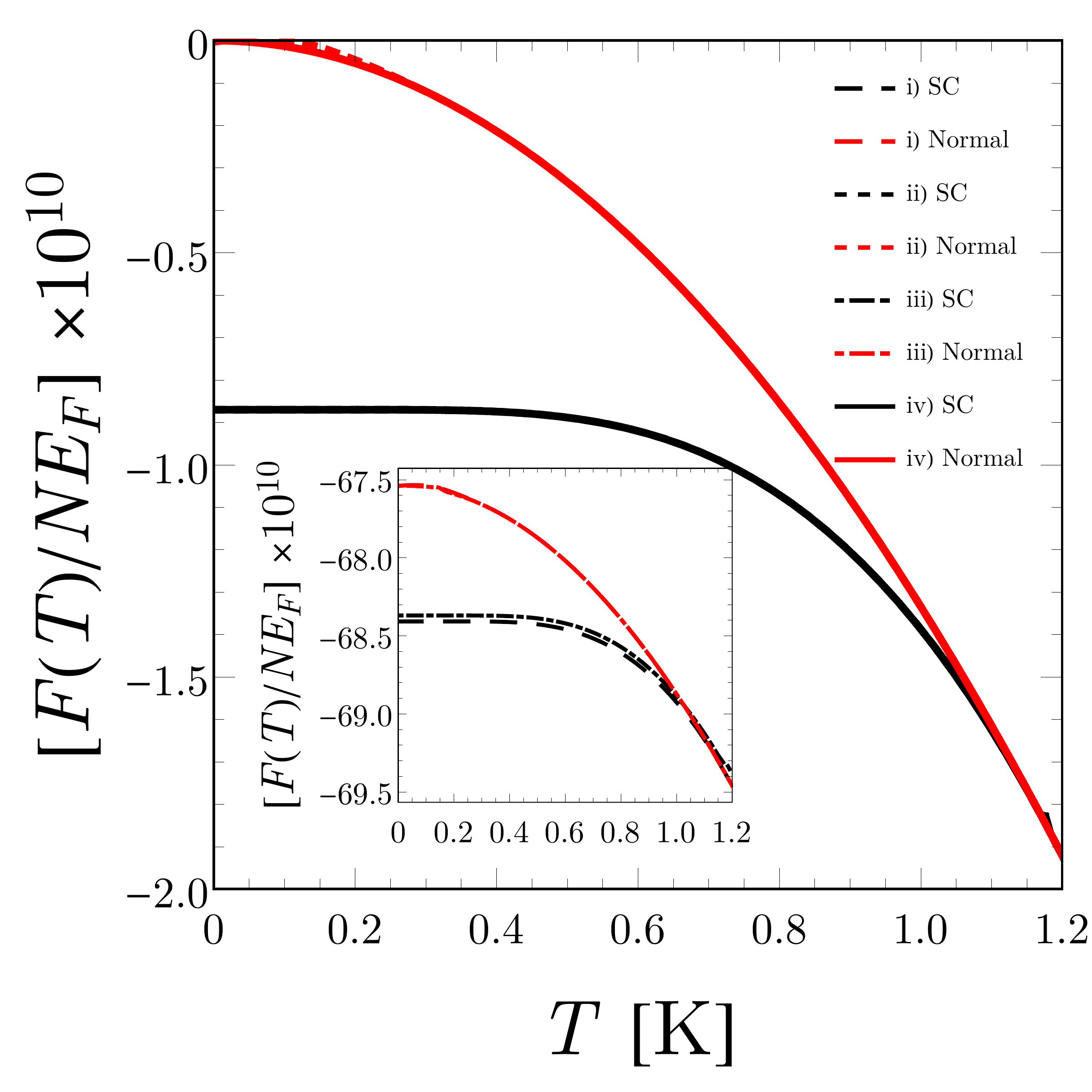}
\vspace*{-0.5cm}
\caption{(Color online) Helmholtz free energy $F(T)/NE_F$ vs. $T$ [K] for aluminum SC for the four cases cited in text. SC free energy (black) curves are below the normal free energy (red) curves as expected. All curves are referenced to the exact normal free energy at $T=0$. Inset shows the SC and normal, black and red curves, respectively of the Helmholtz free energy for the (i) and (iii) cases. Note they are three orders of magnitude greater than the (ii) and (iv) cases.
}\label{Fig-Al-Helmholtz}
\end{figure}

Fig.~\ref{Fig-Al-Helmholtz}a shows the Helmholtz free energy $F(T)/NE_F$ vs. $T$ [K] for all approximation cases previously mentioned. All SC cases behave as expected, i.e., below the normal free energy, and the order of magnitude corresponds to the condensation energy. Note that cases (i) and (iii) have an order of magnitude greater than the (ii) and exact cases. However, the difference between SC and normal free energy for all cases lead to the condensation energy values of Table~\ref{AlTable-1}.
\onecolumngrid\

\twocolumngrid\
\FloatBarrier


\begin{thebibliography}{99}
%
\bibitem{BCS} J. Bardeen, L.N. Cooper, and J.R. Schrieffer, Phys. Rev. \textbf{108}, 1175 (1957)
%
\bibitem{cooper} L.N. Cooper, Phys. Rev. \textbf{104}, 1189 (1956)
%
\bibitem{bardeen55} J. Bardeen and D. Pines, Phys. Rev. \textbf{99}, 1140 (1955) 

\bibitem{morel-anderson} P. Morel and P.W. Anderson, Phys. Rev. \textbf{125}, 1263 (1962)
%
\bibitem{ashcroft} N.W. Ashcroft and N.D. Mermin, \textit{Solid State Physics%
} (Saunders College Publishing, USA, 1976)

\bibitem{GBEC1} V.V. Tolmachev, Phys. Lett. A \textbf{266}, 400 (2000)
%
\bibitem{GBEC2} M. de Llano and V.V. Tolmachev, Physica A \textbf{317}, 546
(2003)
%
\bibitem{bogo58} N.N. Bogoliubov, JETP \textbf{34}, 41 (1958)
%
\bibitem{valatin} J.G. Valatin, N. Cim. \textbf{7}, 843 (1958)
%
\bibitem{GBEC3} M. de Llano and V.V. Tolmachev, Ukr. Phys. J. \textbf{55},
79 (2010)
%
\bibitem{chavez17} I. Ch\'{a}vez, L.A. Garc\'{\i}a, M. Grether and M. de
Llano, Int. J. Mod. Phys. B \textbf{31}, 1745013 (2017)
%
\bibitem{chavez18} I. Ch\'{a}vez, L.A. Garc\'{\i}a, M. Grether, M. de Llano
and V.V. Tolmachev, J. Supercond. Nov. Magn. \textbf{31}, 631 (2018)
%
\bibitem{fetter} A.L. Fetter and J.D. Walecka, \textit{Quantum theory of many-particle systems} (Dover Publications, 2003) 
%
\bibitem{vandermarel} D. van der Marel, Physica C \textbf{165}, 35 (1990)
%
\bibitem{chavez22} I. Chávez, P. Salas, O.A. Rodr\'{i}guez, M. de Llano and M.A. Solís, \textit{Chemical potential influence on the condensation energy of a superconductor in a boson-fermion model of superconductivity}, Physica C (accepted)

\bibitem{kittel} C. Kittel, \textit{Introduction to Solid State Physics} (Wiley 2005) p. 267

\bibitem{harris} E.P. Harris and D.E. Mapother, Phys. Rev. \textbf{165}, 522 (1968)

\bibitem{shoenberg} D. Shoenberg, \textit{Superconductivity} (Cambridge University Press, Cambridge, UK, 1952), 2nd ed.

\bibitem{roberts} B.W. Roberts, J. Phys. Chem. Ref. Data \textbf{5}, 581 (1976)

\bibitem{biondi} M.A. Biondi and M.P. Garfunkel, Phys. Rev. \textbf{116}, 853 (1959)

\bibitem{poole} C.P. Poole, Jr. \textit{et al}.,\textit{ Superconductivity} (Academic Press,
Elsevier, New York, 2007).
%
\end{thebibliography}
\end{document}